\documentclass[twocolumn,showpacs,preprintnumbers,amsmath,amssymb]{revtex4}
\usepackage{amsmath}
\usepackage{amssymb}
\usepackage{txfonts}

\usepackage{graphicx}
\usepackage{dcolumn}
\usepackage{bm}
\usepackage{color}

\begin{document}



\title{Cooperative
atomic scattering of light from a laser with a colored noise
spectrum }

\author{Xiaoji Zhou}\thanks{Electronic address: xjzhou@pku.edu.cn }
\affiliation{School of Electronics Engineering $\&$ Computer
Science, Peking University, Beijing 100871, China}
\date{\today}

\begin{abstract}
The collective atomic recoil lasing is studied for an ultra-cold and
collisionless atomic gas in a partially coherent pump with a colored
noise.  Compared to white noise, correlations in colored noise are
found to be able to greatly enhance or suppress the growth rate,
above or below a critical detuning. Effects on cooperative
scattering of light for noise correlation time, noise intensity and
pump-probe detuning are discussed.  This result is consistent with
our simulation and linear analysis about the evolution equations in
the regions of instability.
\end{abstract}

\pacs{42.60.Mi,05.45.Xt, 42.65.Sf, 42.50.Wk }

\maketitle
\section{Introduction}
Collective nonlinear interactions between cold atoms and light have
attracted considerable attention since collective atomic recoil
lasing (CARL) was observed in cold atoms~\cite{Courteille,Slama1}
and superradiance from Bose-Einstein Condensation (BEC) was realized
in experiments~\cite{Inouyescience}. The signature of CARL is sudden
buildup in the probe field or back-scattering beam oriented
reversely to a coherent pump strongly interacting with an atomic
gas~\cite{Bonifacio, Robb1, Kruse1}. This self-organization
phenomenon shows spontaneous formation of an atomic density grating,
and offers the possibility to study the light amplification derived
from collective interaction of light with cold atomic gases
~\cite{Franca, Rob,Piovella,Piovella1}.

Most studies so far have focused on the coherent laser pump,
neglecting the fact that environment is intrinsically noisy. As
found in a recent work for a partially coherent pump field with a
white noise spectrum~\cite{robb}, the probe intensity can be larger
than that in a coherent pump due to the noise.  The intensity
fluctuation and noise play important roles in this
nonlocally-coupled many-body system. Although the white-noise
assumption is convenient for mathematical treatment, it is somewhat
unrealistic, because fluctuations in the microscopic dynamics have a
finite (nonzero) correlation time, commonly referred to as colored
noise~\cite{Dixit}. The white noise approximation is valid when the
correlation time of fluctuation is much shorter than all other
relevant time scales in this problem.  In other cases there are
discrepancies between white-noise theory and experiments, such as
the photon statistics of a dye laser output where the relative
intensity fluctuations tend to increase indefinitely as the laser is
weakly excited~\cite{Dixit, Short,Cao}, the reversed asymmetry in
the doublet spectrum of double optical resonance is reverted back to
normal  for lager detuning~\cite{Zoller}. Hence, the colored noise
model is more realistic than the white-noise model and is widely
studied for laser systems, for example in the sideband squeezing in
intracavity second-harmonic generation~\cite{Walls}, excess quantum
noise in a laser~\cite{Lee}, and four-wave mixing~\cite{Agarwal}. In
these systems, the colored noise is caused by the laser frequency
fluctuation, and the field line shape is Gaussian in some parameter
regime, different with the Lorentzian line shape for white noise.

In this paper we study CARL based on a laser pump with colored
noise. Effects of noise on the cooperative scattering of light are
studied by stochastic simulation and linear analysis at small
correlation time. The Lorentzian bandwidth is related to noise
intensity, pump-probe detuning, and the cut-off frequency given by
the inverse of correlation time.  Effects of these parameters on the
growth rate of amplification are analyzed and discussed.

\section{model}
We consider the normal setup for CARL~\cite{Courteille,robb,Bon}.
The backscattered light named probe beam has amplitude $E_{1}$ and
frequency $\omega_{1}$, and a partially coherent strong pump field
has amplitude $E_{2}$ and the mean frequency $\omega_{2}$. These two
beams with almost the same frequency $|\omega_{1}-\omega_{2}|\ll
\omega_{1}$ or $ \omega_{2}$, are approximately counterpropagating
to form a spatially periodic optical lattice potential, interacting
with a very cold and collisionless atomic gas. Atoms moving in this
lattice form a density grating, and in turn play a role on the
evolution of probe field.

Since $|\omega_{2}-\omega_{1}|$ is assumed to be very small, the
pump or probe wave vector is given by $k=2 \pi/\lambda$ where
$\lambda \equiv \lambda_{2} \approx \lambda_{1}$.  The pump
frequency $\omega \equiv \omega_2$ is detuned from the atomic
resonance frequency $\omega_{0}$ by $\Delta=\omega-\omega_{0}$. The
two-photon recoil frequency is given by $\omega_{r}=2\hbar k^2/m$.
In terms of the rescaled time variable $\tau=\omega_{r}\rho t$, the
rescaled amplitude of probe field $A=\sqrt{\frac{2
\varepsilon_{0}}{n\hbar\omega \rho}} E_{1}$, the rescaled position
$\theta_{j}=2 k z_{j}$ and momentum $p_{j}=(mv_{j})/(\hbar k \rho)$
of the jth atom, the modified classical CARL equations with a
stochastic pump phase noise are given by~\cite{robb}
\begin{eqnarray}
 \label{1}    d\theta_{j}/d\tau&=&p_{j},\\
     dp_{j}/d\tau&=&-A e^{i(\theta_{j}-\phi)}+c.c.,\\
  \label{3}  dA/d\tau&=&-\langle\langle e^{-i(\theta_{j}-\phi)}\rangle\rangle +i\delta A,\\
\label{4}     d\phi/d\tau&=&\varepsilon(\tau),
\end{eqnarray}
where the dimensionless CARL parameter is given by
$\rho=(\frac{\Omega g \sqrt{n}}{2\Delta \omega_{r}})^{2/3}$, the
pump Rabi frequency is $\Omega=dE_{2}/\hbar$, the atomic density is
$n$, the atom-mode coupling constant is given by $g=d
\sqrt{\frac{\omega }{2 \varepsilon_{0} \hbar}}$, and the dipole
matrix element for the atomic transition is $d$.  The average in
Eq.(\ref{3}) is defined as the average over all atoms,
$$\langle\langle \cdots \rangle\rangle\equiv \frac{1}{N}\times
\sum_{j=1}^{N} (\cdots)_{j}.$$ and the scaled pump-probe detuning is
given by $\delta=(\omega_{2}-\omega_{1})/(\omega_{r}\rho)$.

The partial coherence of pump field is described by a phase
diffusion model, assumed to evolve according to Eq.(\ref{4}), where
$\varepsilon(\tau)$ is a Gaussian random variable with zero mean
$\langle\varepsilon(\tau)\rangle\equiv
\overline{\varepsilon(\tau)}=0$, and
variance~\cite{Dixit,Agarwal,Honeycutt}
\begin{eqnarray}\label{form}
\langle\varepsilon(\tau)\varepsilon(\tau')\rangle\equiv
\overline{\varepsilon(\tau)\varepsilon(\tau')}
=\frac{\Gamma}{\tau_{0}}
     e^{-\frac{|\tau-\tau'|}{\tau_0}}.
\end{eqnarray}
The colored laser noise given by Eq.(5) is parameterized by the
noise correlation time $\tau_{0}$ and  the noise strength $\Gamma$,
where $\langle \cdots \rangle$ denotes the average over sampling. In
the limit $\tau_0\rightarrow 0$, from Eq. (5) we get $\langle
\xi(\tau)\xi(\tau')\rangle \rightarrow 2\Gamma\delta(\tau-\tau')$,
corresponding to the case of a white noise in laser phase
fluctuations with a Lorentzian line shape and linewidth [half width
at half maximum (HWHM)] $\Gamma$ for the pump field~\cite{robb}.
When $\tau_0\rightarrow\infty$, it recovers the case with a coherent
pump laser~\cite{Courteille,robb}. For $1/\tau_{0}\gg \Gamma$, the
field line shape is a Lorentzian with a full width at half maximun
(FWHM) of $2\Gamma$, while for $1/\tau_{0}\ll \Gamma$ it is a
Gaussian with a FWHM related to $\sqrt{\Gamma/\tau_{0}}$.  The
general values of $\Gamma \tau_{0}$ between these two limits lead to
Voigt profiles~\cite{Agarwal}, meaning that the laser spectrum is
significantly non-Lorentzian for detuning greater than $1/\tau_{0}$
(from line center) and the spectrum cuts off much more rapidly than
a Lorentzian.  Consequently $1/\tau_{0}$ is known as the cut-off
~\cite{Zoller}.  Eq.(5) can also be written as
$\dot{\varepsilon}(\tau)+\frac{1}{\tau_{0}}\varepsilon(\tau)=\frac{1}{\tau_{0}}
\sqrt{2 \Gamma} \varsigma ({\tau})$, where $\varsigma ({\tau})$ is
the Gaussian white noise satisfying $\langle \varsigma
({\tau})\rangle=0$, $\langle \varsigma
({\tau})\varsigma({\tau'})\rangle=\delta(\tau-\tau')$~\cite{Walls}.
As pointed out in the Ref.~\cite{robb}, the amplification due to the
partially coherent pump is less sensitive to the atomic distribution
in the form of $\exp(-\frac{p_{0}^{2}}{2\sigma^{2}})$ than that due
to a coherent pump, where the scaled initial momentum is $p_{0}$,
$\sigma=\frac{\sqrt{3m\kappa_{B}T}}{\hbar k \rho}$, $\kappa_{B}$ is
Boltzmann's constant, and $T$ is the temperature of atomic gas. In
this paper we consider effects of correlation time while assuming
$\sigma=0$.

\section{Stochastic simulation}
\begin{figure}[tbp]
   \begin{center}
        \includegraphics[height=7cm]{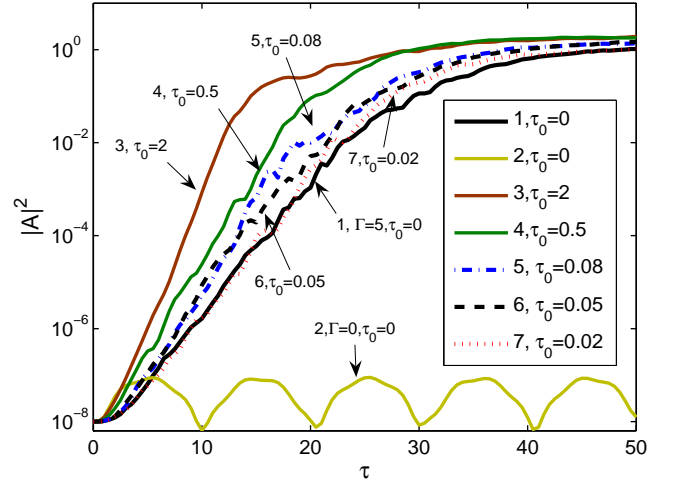}
    \end{center}
  \caption{(Color online) The scaled probe intensity $|A|^{2}$ (averaged over 100 runs) for
  a cold gas ($\sigma=0$) by a partially coherent pump with detuning $\delta=2$ for
  different correlation time $\tau_{0}$. The solid line 2 is for noise intensity $\Gamma=0$, while
  all other lines are with $\Gamma=5$. }
  \label{fig:1}
\end{figure}

We directly solve the set of stochastic ordinary differential
equations (\ref{1})-(\ref{4}) using stochastic Runge-Kutta
algorithms with a colored noise~\cite{Honeycutt}. To conveniently
compare with the case of a white noise, in the following we use the
same value $\Gamma=5$ and detuning $\delta=2,5,10$, as in
Ref.~\cite{robb}. The scaled probe intensity over time is plotted in
Fig.\ref{fig:1} to show effects of pump phase diffusion for
pump-probe detuning $\delta =2$. The solid line 1 is for the case
with white noise $\Gamma=5$ and $\tau_{0}=0$ ~\cite{robb}. The case
of coherent pump with $\tau_{0}=0$ and $\Gamma=0$ is shown in the
solid line 2, where the intensity of probe beam is very low and
oscillates with time, and the gain is suppressed. The growth rate
and intensity of the backscattered field greatly increase under the
partially coherent pump. In the case of colored noise, the growth
rates and amplitudes of the back scattering beams are increased
compared to the case of white noise, and this gain increases with
correlation time $\tau_{0}$, as shown in the dotted line
($\tau_{0}=0.02$), dashed line ($\tau_{0}=0.05$), dash-dotted line
($\tau_{0}=0.08$), solid line 4 ($\tau_{0}=0.5$), and solid line 3
($\tau_{0}=2$) for $\delta=2$.
\begin{figure}
  \includegraphics[height=7cm]{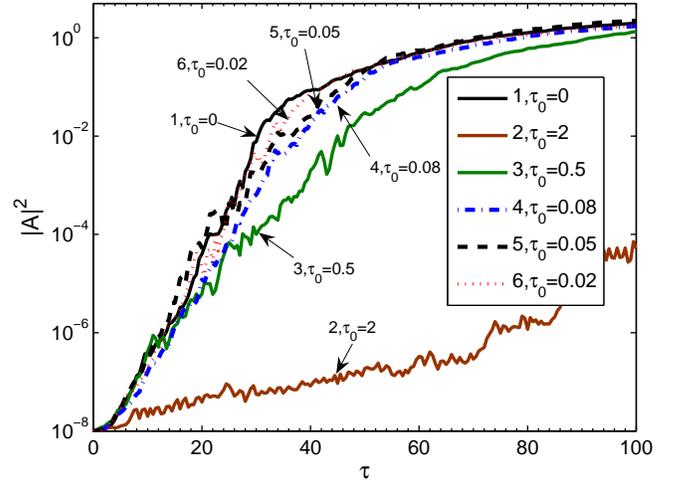}
  \caption{(Color online) The scaled probe intensity $|A|^{2}$
  versus time for  detuning $\delta=5$,$\Gamma=5$.}
  \label{fig:2}
\end{figure}
The correlation time of noise has an opposite effect with $\delta
=5$, as shown in Fig.\ref{fig:2}. The growth rates become slow, and
it take more time to get to the saturation than in the case of
$\delta=2$. The growth rate and intensity of probe field is
suppressed for $\tau_{0}=2$, as shown in the solid line 2. For small
correlation time $\tau_{0}=0.02$(dotted line), it coincides with the
case of white noise (the solid line 1 $\tau_{0}=0$). The growth rate
decreases with the increase in correlation time, as shown in the
dashed line ($\tau_{0}=0.05$), and dash-dotted line
($\tau_{0}=0.08$). However, the difference between the colored noise
and white noise at these small correlation times is not so obvious
for $\delta=5$. For $\tau_{0}=0.5$, there is a distinct decrease in
the growth rate and intensity of probe beam comparing to the case of
white noise, as shown in the solid line 3. When the detuning is
further increased to $\delta=10$, as shown in Fig.\ref{fig:3}, the
rate and intensity of gain are greatly suppressed for correlation
time $\tau_{0}=2$ (solid line 2) and $\tau_{0}=0.5$ (solid line 3).
At  small $\tau_{0}$,
 growth rates also decrease with the increase in correlation time, as
shown by the dotted line ($\tau_{0}=0.02$), dashed line
($\tau_{0}=0.05$) and dash-dotted line ($\tau_{0}=0.08$).

The above results show that the gain is suppressed with the increase
in correlation time for large detuning ($\delta=5$ and $\delta=10$).
The condition of gain is destroyed with the long correlation time of
noise.  However for the case with small detuning $\delta=2$, the
long correlation time helps to enhance the amplification in the
growth rate and amplitude of the probe beam. According to our
simulation, there are similar results for $\Gamma=3$ and $\Gamma=1$,
where a long correlation time is able to suppress the growth rate
for $\delta=5$ and $\delta=10$, and enhance it at $\delta=2$. Thus
depending on detuning, the increase in correlation time can either
enhance or suppress the gain. To further understand these different
behaviors, we carry out the linear analysis at small correlation
time in the next section.
\begin{figure}
  \includegraphics[height=7cm]{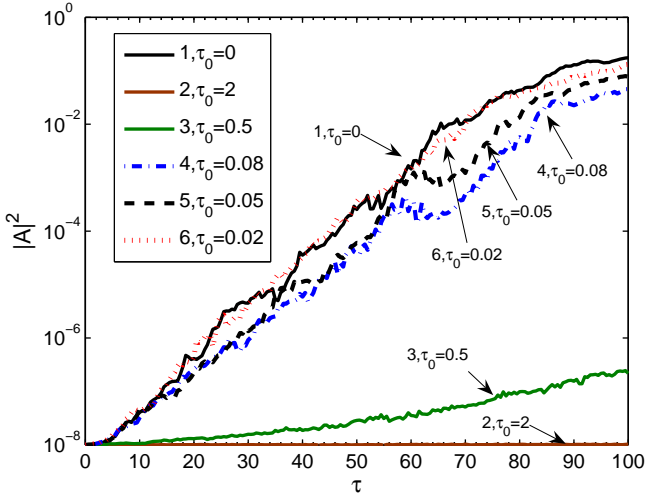}
   \caption{(Color online) The scaled probe intensity $|A|^{2}$ versus  time for detuning
   $\delta=10$,$\Gamma=5$.}
 \label{fig:3}
\end{figure}

\section{linear analysis}
To analyze effects of  colored noise in the regions of instability,
we derive the linear evolution equations for the average
scattered-field intensity $ \overline{A^{\ast} A}$ from equations
Eqs.(\ref{1})-(\ref{4}),
\begin{eqnarray}
\label{asqu}    \frac{d\overline{|A|^{2}}}{d\tau}&=&
\overline{A^{\ast} b}+c.c.,\\
\label{ab} \frac{d \overline{A^{\ast} b }}{d\tau}&=&-i
\overline{A^{\ast} P}+\overline{b^{\ast} b}-(i\delta+\Gamma')
\overline{A^{\ast} b},\\
\frac{d \overline{A^{\ast} P} }{d\tau}&=&-\overline{|A|^{2}}+\overline{b^{\ast} P}-(i\delta+ \Gamma') \overline{A^{\ast} P},\\
\frac{d \overline {|b|^2}}{d\tau}&=& -i \overline {b^{\ast} P}+c.c.,\\
\frac{d \overline {b^\ast P}}{d\tau}&=& -\overline{A b^{\ast}} +i \overline{{|P|}^{2}},\\
 \label{psqu}     \frac{d \overline {|P|^{2}}}{d\tau}&=& -\overline{A^{\ast} P}
 +c.c.,
\end{eqnarray}
where $\overline{|A|^{2}}=\overline{A^{\ast} A}$,
$\overline{|b|^{2}}=\overline{b^{\ast} b}$,
$\overline{|P|^{2}}=\overline{P^{\ast} P}$, and
\begin{equation}
    b=\frac{1}{N}\sum_{j}^{N}e^{-i(\theta_{j}-\phi)},P=\frac{1}{N}\sum_{j}^{N}(p_{j}e^{-i(\theta_{j}-\phi)}).
\end{equation}

To understand how to derive the above formula, here we show the
steps in getting Eq.(\ref{ab}).  Other equations can be obtained
with similar steps.  From Eq.(\ref{3}), we have
\begin{equation}
\frac{d \overline{A^{\ast} b }}{d\tau}=-i \overline{A^{\ast}
P}+\overline{b^{\ast} b}-i \delta \overline{A^{\ast} b}
 + i \overline{\varepsilon A^{\ast} b},
\end{equation}
where the last term can be expressed as
\begin{equation}
 i \overline{\varepsilon A^{\ast} b}=i\overline{\frac{A^{\ast}}{N}\sum_{j=1}^{N} e^{-i\theta_{j}}
 e^{i\int_{T}^{\tau_{1}}
  \varepsilon(\tau')d\tau'} f(\tau_{1},T)},
\end{equation}
with
\begin{equation}\label{f}
f(\tau_{1},T)\equiv \langle \varepsilon(\tau)e^{\int_{\tau_{1}}^{T}
i \varepsilon(\tau')d\tau'}\rangle.
\end{equation}
We can expand Eq.(\ref{f}) with the series
\begin{eqnarray*}
f(\tau_{1},T)&=&\sum_{m=0}^{\infty}\frac{1}{2^{m}m!}\int_{\tau_{1}}^{T}
i \langle \varepsilon(\tau_{\mu})\varepsilon(\tau_{1})\rangle
d\tau_{\mu}\\&\times&(\int_{\tau_{1}}^{T}
d\tau_{\mu_{1}}\int_{\tau_{1}}^{T}  d\tau_{\mu_{2}} i^{2}\langle
\varepsilon(\tau_{\mu_{1}})\varepsilon(\tau_{\mu_{2}})\rangle)^{m}.
\end{eqnarray*}
For small correlation time $\tau_{0}$, it can be further expressed
for the moments $m=0$ and $m=1$ only, $$ f(\tau_{1},T) \approx
i\Gamma(1- e^{-\frac{T-\tau_{1}}{\tau_{0}}})[1+\frac{\Gamma
 \tau_{0}(1-e^{-\frac{T-\tau_{1}}{\tau_{0}}})}{2}
 -\frac{\Gamma(T-\tau_{1})}{2}].$$
Assuming $T=\tau_{1}+K \tau_{0}$ where $K$ is an adjustable
parameter, $1\ll K\ll(T-\tau_{1}$), we have
\begin{eqnarray}\label{ab1}
 i \overline{\varepsilon A^{\ast} b}=-\Gamma' \overline{A^{\ast} b}
\end{eqnarray}
with
\begin{equation}
\Gamma'=\Gamma[1-(K-1)\Gamma \tau_{0}/2].
\end{equation}

\begin{figure}[tbp]
\begin{center}
     \includegraphics[height=6 cm]{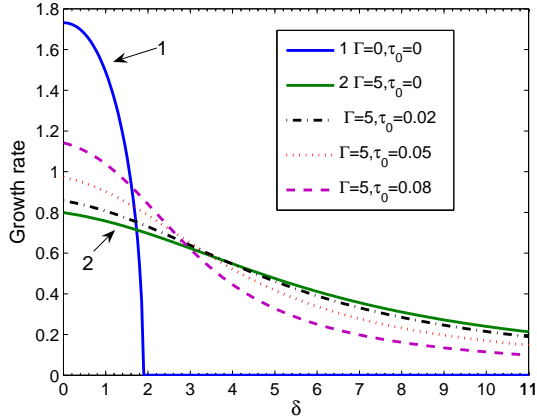}
    \end{center}
  \caption{The growth rate of probe beam $\Re e
(\lambda)$ versus detuning $\delta$
  for different correlation times. The solid line 1 is for
the coherent pump with
  $\Gamma=0,\tau_{0}=0$, and the solid line 2 is for the partially coherent pump with white noise
  $\Gamma=5,\tau_{0}=0$.  Other lines are for the partially coherent pump with colored noise intensity $\Gamma=5 $ and
  $K=4$, and with correlation time $\tau_{0}=0.08$(dashed line), $\tau_{0}=0.05$ (dotted line), or $\tau_{0}=0.02$ (dash dotted line ).}
  \label{fig:4}
\end{figure}

Eqs. (\ref{asqu}-\ref{psqu}) are identical with equations (5)-(10)
in Ref.~\cite{robb} except that $\Gamma$ for the white noise is now
replaced by $\Gamma'$ for the colored noise. Thus we can use the
relation between the white noise and colored noise models given by
Eq.(17) for small $\tau_{0}$.  An increase of $\tau_{0}$ can be
regarded as decreasing the noise intensity or the linewidth of white
noise. Hence, for the initially cold and collisionless atomic gas,
we obtain the characteristic root $\lambda$ from the solution of
Eqs. (\ref{asqu}-\ref{psqu}). The region of instability or
amplification satisfies $\Re e (\lambda)>0$. Based on the solution,
the growth rates of the probe beams $\Re e (\lambda)$ as the
detuning $\delta$ for the different correlation time are plotted in
Fig.\ref{fig:4}.

As shown in Fig.\ref{fig:4}, the maximum growth rate occurs at
$\delta=0$. In the case of a coherent pump shown by the solid line
1, the cutoff is so sharp that no instability or amplification of
the probe beam occurs above $\delta_{c}=(27/4)^{1/3}\approx 1.9$.
The noise can extend regions of amplification because the pump phase
diffusion broadens the region of instability, as shown by the solid
line 2 for the partially coherent pump with a white noise
$\Gamma=5,\tau_{0}=0$. The case with small $\tau_{0}=0.02$ is very
close to the case of white noise~\cite{robb}. For detuning
$\delta=2$, the growth rates increase with increasing $\tau_{0}$, as
shown in Fig.4. However, as the detuning increases, the growth rates
decrease with the increase in  correlation time for $\delta=5$.
Compared to the case of white noise, for example, the growth rate is
increased below the critical value $\delta_{0}\approx 3.3$, and
decreased  above it with the correlation time $\tau_{0}=0.05$, as
shown in Fig.4.  Hence, depending on detuning, the effect on the
growing rate is different for different correlation times under the
same noise intensity. For the different noise strength, the growth
rates versus  detuning with $\tau_{0}=0.05$ are drawn in Fig.5. We
got $\delta_{0}\approx 1.8$ for $\Gamma=1$, $\delta_{0}\approx 2.7$
for $\Gamma=3$. The critical detuning $\delta_{0}$ decreases with
the decrease in noise strength $\Gamma$. The growth rates are almost
same for  white noise and colored noise with small noise intensity
$\Gamma=1$, and there is an obvious difference for $\Gamma=3$.  The
increase in the correlation time is beneficial to the growth rate of
the amplification near the threshold detuning, while destructive for
the big detuning.

\begin{figure}[tbp]
\begin{center}
     \includegraphics[height=6 cm]{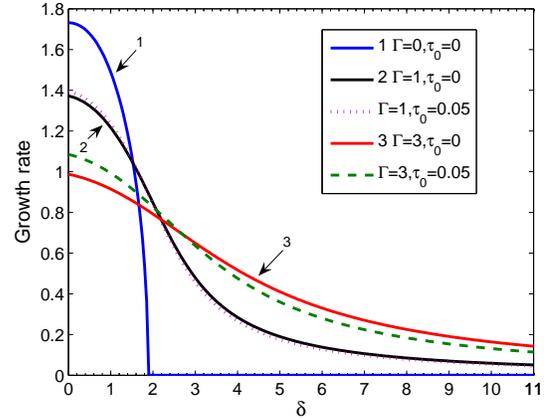}
    \end{center}
  \caption{The growth rate of the probe beam $\Re e
(\lambda)$ versus detuning $\delta$
 for different noise strength. The solid line 1 is for the
coherent pump with
  $\Gamma=0,\tau_{0}=0$. The solid line 2 is for white noise with
  $\Gamma=1,\tau_{0}=0$, and dotted line is for colore noise with $\Gamma=1,\tau_{0}=0.05$.
  The solid line 3 is for $\Gamma=3,\tau_{0}=0$, and dashed line is for
  $\Gamma=3,\tau_{0}=0.05$.}
  \label{fig:5}
\end{figure}

\section{Discussion and conclusions}

The gain of CARL given by the growth rate of the probe beam depends
on the frequency of the light field.  It has a certain gain
bandwidth defined as the spectral range where the light scattering
is exponentially amplified.  In the amplification process, atom
scatter photons from the pump field into the probe beam, experience
an acceleration due to the photonic recoil, and occupy different
momentum states. On the other hand, the frequency of scattered
photons is shifted with respect to the pump-light frequency,
resulting in  pump-probe detuning. In the case of a noisy pump,
there are two velocities involved. One is the stochastic phase
velocity with mean value
$(\omega_{1}-\omega_{2})c/(\omega_1+\omega_{2})$ of the optical
potential, and the other is the atomic velocity.  Due to the
interaction between the atoms and the optical field, the
synchronization between these two velocities eventually leads to the
scaled momenta distribution of atoms around a mean value $\langle p
\rangle\approx -\delta$, when $\delta>\delta_{c}=1.9$ and the
dynamical phase evolution of probe field is negligible.  Thus the
region of instability extends from the threshold value of the
coherent pump $\delta_{c}$ to the bigger detuning~\cite{robb}. At
the same time, a decrease in $\langle p \rangle$ causes an increase
in $|A|^{2}$, due to the conservation of momentum $\langle p
\rangle+|A|^{2}=const$. Hence the saturation of instability occurs
when the scaled probe intensity satisfies $|A|^{2}\approx \delta$.
This is the reason of amplification in large detuning for the
partially coherent pump.

However, the noise correlation time $\tau_{0}$, noise intensity
$\Gamma$, and pump-probe detuning $\delta>\delta_{c}$ greatly affect
the growth rate of this amplification behavior.  To understand this
relation, we define $\gamma$ as the ratio of the effective gain
bandwidth $\Gamma_{eff}$ determined by noise intensity and
correlation time to the detuning,
 \begin{equation}
\gamma=\Gamma_{eff}/\delta.
\end{equation}
It determines how many momentum states lying within the effective
linewidth may participate and be amplified in the CARL dynamics. If
this ratio is one, it means that all the momentum states within the
linewidth are amplified at the same time and the growth rate is the
fastest. If the ratio is more or less than one, this growth rate
decreases.

In the case with small correlation times, $\Gamma\ll 1/\tau_{0}$,
the cut-off of the laser spectrum is larger than Lorentzian
bandwidth. The colored noise model and white noise model lead to
similar results following the relation given by Eq.(17),
$\Gamma_{eff}=\Gamma[1-(K-1)\Gamma \tau_{0}/2]$.  The effective
bandwidth of noise $\Gamma_{eff}$ limits the range of frequencies
accessible for the probe light field. For $\Gamma=5$ and $\delta=2$,
$\gamma\approx 2.5-18.75 \tau_{0}$, the increase in correlation time
leads to the decrease in $\gamma$ closer to one. Hence the growth
rate of amplification is enhanced with increasing $\tau_{0}$, as
shown in Fig.4 of linear analysis and in Fig.1 of simulation results
with $\tau_{0}=0.02,0.05,0.08$. However, for $\delta=5$,
$\gamma\approx1-7.5 \tau_{0}$, the increase in correlation time
leads to a deviation from the match condition, hence increasing
$\tau_{0}$ decreases the growth rate, as shown in Fig.4 of the
linear analysis and in Fig.2 of simulation results. Furthermore, for
$\delta=10$, $\gamma\approx0.5-3.75\tau_{0}$, increasing correlation
time destroys the match condition $\gamma = 1$, as demonstrated in
Fig.4 of linear analysis and in Fig.3 of the simulation results with
$\tau_{0}=0.02,0.05,0.08$ . Based on linear analysis shown in Fig.4,
the simulation results at small correlation times shown in
Fig.(1-3), and the above analysis, we reach the conclusion that the
change in the correlation time can be regarded as adjusting the
effective noise intensity. Whether the change in correlation time
enhances or suppresses the growing rate depends on whether it helps
or destroys the match condition $\gamma=1$.

The large correlation time corresponds to slow frequency fluctuation
$1/\tau_{0}\ll \Gamma$, when the laser shape approaches a Gaussian
with the effective bandwidth of HWHM
$\Gamma_{eff}=\sqrt{[8(ln2)\Gamma/\tau_{0}]}/2$.  The effective
bandwidth limits the range of frequencies accessible for  probe
beam.  When $\delta=2$, we have $\gamma=1.86,0.93$ for
$\tau_{0}=0.5$ and $2$. Because 0.93 is closer to unity than 1.86,
the growth rate at $\tau_{0}=2$ is faster than that at
$\tau_{0}=0.5$, as shown in the solid line 3 and 4 of Fig.1.
However, for $\delta=5$, $\tau_{0}=0.5,2$ corresponding to
$\gamma=0.74,0.37$, the increase of correlation time destroys the
match condition and the amplification at $\tau_{0}=2$ is almost
suppressed, as shown in the sold line 2 of Fig.2.  Furthermore, at
$\delta=10$ and $\tau_{0}=0.5,2$,  $\gamma=0.37,0.18$, these two
values are far from one and the amplification is also suppressed at
$\tau_{0}=0.5$, as shown in the solid line 3 of Fig.3. The
simulation results show that the growth rate is greatly suppressed
with large correlation times. Hence, for big $\tau_{0}$, the
increase in  correlation time greatly reduces the bandwidth, and
suppresses the number of momentum states of participating in the
CARL dynamics. The above explanation also is suitable for
$\Gamma=3$, while it is not right for weak noise intensity
$\Gamma=1$.

It should be possible to observe these effects following the
experimental observation of CARL~\cite{Courteille,Slama1} where
ultracold $^{87}Rb$ atoms were enclosed in a ring cavity.  The
difference between the experimental setup and our model is that the
pump field in the experiment is also in a cavity mode which
counterpropagates with respect to the probe beam. The experimental
reported characteristic growth time $t_{g}$ for the instability of
CARL is about $1\mu s$~\cite{Kruse1}, corresponding to
$\omega_{r}\rho\approx t_{g}^{-1}=10^{6} s^{-1}$, $\rho\approx 10$,
and the recoil shift of Rb as $\omega_{r}=2 \hbar k^{2}/m\approx 2
\pi \times 14 k Hz$. For a high finesse of the cavity $\delta
\omega_{pump}\approx 2 \pi \times 20 kHz$ and the scaled pump
linewidth $\Gamma= \delta \omega_{pump}/(\omega_{r}\rho)\approx
0.02$, it corresponds to the case of a coherent pump field for
$\Gamma \ll 1$. However, the linewidth of the pump field can be
adjusted between good-cavity and bad-cavity regimes by varying the
finesse of cavity, atom number, and pump power.  $\Gamma$ from 0.3
to 0.2 corresponding to $\rho=4.7$ to $7.0$ is the semiclassical
good-cavity regime. $\Gamma$ from $3.7$ to $2.8$ corresponding to
the parameter $\rho=5.1$ to $6.7$ is the typical semiclassical
bad-cavity regime~\cite{Courteille,Slama1}. If we assume $\rho=5$
for Rb, the scaled noise intensity $\Gamma=5$ means a linewidth of
the Lorentzian of $2\pi\times 350 k Hz $, the scaled correlation
time $\tau_{0}=0.5$ means the cutoff at frequency $2\pi\times 140 k
Hz$, and pump-probe detuning $\delta=2,5,10$ corresponds to $140
kHz, 350 kHz, 700kHz$, respectively.  Hence, to certain extent, our
analysis can be regarded as the CARL in  bad-cavity regime which is
possible to be accessed in experiments.

In conclusion, the correlation time of pump phase noise greatly
affects the growth rate and intensity of cooperative scattering in
the system of CARL.  The noise makes the amplification region for
pump-probe detuning larger than that in the coherent pump. The noise
intensity and correlation time determine the effective linewidth of
pump laser, not just noise intensity in the case of white noise
predicted by phase diffusion model of ideal laser theory. The change
in correlation time can enhance or suppress the growing rate
depending on the ratio of the effective bandwith to pump-probe
detuning  which determines how many momentum states within the
linewidth are amplified at the same time. This ratio equal to one
corresponds to the best match condition for big noise intensity.
Whether the growth rate is suppressed or enhanced by the change in
correlation time depends on whether it helps or destroy the
synchronization condition.  These results are useful for analyzing
the cooperative scattering process and effects of noise on the
collective nonlinear interaction between cold matter and light, and
also helpful in studying superradiance from BEC~\cite{2003Theo} or
the phase coherent matter-wave amplification~\cite{Kozuma} because
of the same gain mechanism.

I am grateful to L. Cao and D. J. Wu for their help in the
calculation, G. R. M. Robb and X. Xu for their help in the
simulation, L. Yin for critical reading our manuscript. I thank two
unknown referees for useful and detailed suggestions. This work is
partially supported by the state Key Development Program for Basic
Research of China (No.2005CB724503, 2006CB921402) and NSFC (No.
60490280).


\begin{references}

\bibitem{Courteille} S. Slama, S. Bex, G. Krenz, C. Zimmermann, Ph.
W. Courteille, \prl {\bf 98} 053603 (2007).

\bibitem{Slama1} S. Slama, G. Krenz, S. Bex, C. Zimmermann, Ph. W. Courteille, \pra {\bf
75} 063620 (2007).

\bibitem{Inouyescience} S. Inouye, A. P. Chikkatur, D. M.
    Stamper-Kurn, J. Stenger, D. E. Pritchard, and W. Ketterle, Science
    {\bf 285}, 571 (1999); D. Schneble, Y. Torii, M. Boyd, E. W. Streed, D. E.
    Pritchard, and W. Ketterle, Science {\bf 300}, 475 (2003).

\bibitem{Bonifacio} R. Bonifacio and L. De Salvo, Nucl. Instrum. Methods
Phys. Res., Sect. A {\bf 341}, 360 (1994); R. Bonifacio, L. De
Salvo, L. M. Narducci, and E. J. D¡¯Angelo, Phys. Rev. A {\bf 50},
1716 (1994).

\bibitem{Robb1} G. R. M. Robb, N. Piovella, A. Ferraro, R. Bonifacio,
Ph. W. Courteille, C. Zimmermann, \pra {\bf 69}, 041403 (2004).
\bibitem{Kruse1} D. Kruse, C. V. Cube,C. Zimmermann, Ph. W. Courteille, \prl {\bf 91}, 183601
(2003).

\bibitem{Franca} V. V. Franca, G. A. Prataviera, \pra {\bf 75}, 043604
(2007).
\bibitem{Rob} G. R. M. Robb, B. W. J. McNeil, Phys. Rev. Lett. {\bf 94}, 023901
(2005).
\bibitem{Piovella}  N. Piovella, M. Cola, R. Bonifacio, Phys. Rev. A {\bf 67},
013817 (2003).
\bibitem{Piovella1} N. Piovella, M. Gatelli, R. Bonifacio, Opt. Commun. {\bf 194}, 167(2001).

\bibitem{robb} G. R. M. Robb and W. J. Firth, Phys. Rev. Lett. {\bf 99},
253601 (2007).

\bibitem{Dixit} S. N. Dixit, P. S. Sahni, Phys. Rev. Lett. {\bf 50},
1273 (1983).
\bibitem{Short} R. Short, L. Mandel, R. Roy, Phys. Rev. Lett. {\bf 49},
647 (1982).
\bibitem{Cao} L. Cao, D. J. Wu, X. L. Luo, \pra {\bf 45}, 57(1993).

\bibitem{Zoller} P. Zoller, P. Lambropoulos, J. Phys. B: Atom.
Molec. Phys. {\bf 12}, L547 (1979).

\bibitem{Walls} T. A. B. Kennedy, T. B. Anderson, D. F. Walls, \pra {\bf 40}, 1385 (1989).


\bibitem{Lee} A. M. Lee, M. P. Exter, A. L. Mieremet, N. J. van
Druten, and J. P. Woerdman  Phys. Rev. Lett. {\bf 81}, 5121 (1998).

\bibitem{Agarwal} G. S. Agarwal, G. Vemuri, C. V. Kunasz, J. Cooper, \pra {\bf
46}, 5879(1992).

\bibitem{Bon} R. Bonifacio, G. R. M. Robb, and B.W. J. McNeil,
Phys. Rev. A {\bf 56}, 912 (1997).

\bibitem{Honeycutt}R. L.Honeycutt, \pra {\bf 45}, 604(1992).


\bibitem{2003Theo}H. Pu, W. P. Zhang, and P. Meystre, Phys. Rev. Lett. {\bf 91},150407
(2003).

\bibitem{Kozuma} M. Kozuma, Y. Suzuki, Y. Torii, T. Sugiura, T. Kuga, E. W. Hagley, L. Deng, Science {\bf 286}, 2309 (1999).









\end{references}
\end{document}